\documentclass[conference]{IEEEtran}
\usepackage{epsfig}

\usepackage{cite}
\usepackage{array}
\usepackage{mdwmath}
\usepackage{mdwtab}
\usepackage{url}

\usepackage[tight,footnotesize]{subfigure}
\usepackage{url}

\begin{document}
\title{Hardware Fingerprinting Using HTML5}

\author{\IEEEauthorblockN{Gabi Nakibly,
Gilad Shelef,
Shiran Yudilevich}
\IEEEauthorblockA{Computer Science Department,\\
Technion -- Israel Institute of Technology.}}

\maketitle

\begin{abstract}
Device fingerprinting over the web has received much attention both by the research community and the commercial market a like. Almost all the fingerprinting features proposed to date depend on software run on the device. All of these features can be changed by the user, thereby thwarting the device's fingerprint. In this position paper we argue that the recent emergence of the HTML5 standard gives rise to a new class of fingerprinting features that are based on the \emph{hardware} of the device. Such features are much harder to mask or change thus provide a higher degree of confidence in the fingerprint. We propose several possible fingerprint methods that allow a HTML5 web application to identify a device's hardware. We also present an initial experiment to fingerprint a device's GPU.   
\end{abstract}

\section{Introduction}
Identifying a user -- without its explicit cooperation -- as it browses the web is a much coveted goal of the websites the user visits. Identifying a visiting user allows a website to combat fraud and detect breaches of the website's terms of use. On the flip side it can be used to compromise the privacy of a user by tracking him and identifying his browsing habits and history. This is most commonly used to serve advertisements tailored to the user. In most cases, identifying a user without his cooperation is reduced to identifying the device it uses to browse the web. The most straightforward method to track a device is by using HTTP cookies. This allows a web site to store state on the user's device, which is then sent back to the website
upon subsequent visits. 

In the last few years more and more cases were observed in which HTTP cookies were abused to track a user and his browsing history~\cite{cookielessmonster}. This has heightened the user community awareness of the privacy threats imposed by HTTP cookies. A recent study~\cite{comscore} in Australia has shown that $1/3$ of users delete cookies within a month after their visit to a website. In addition, there are numerous plug-ins and browser extensions that allow a user to keep tabs on the cookies set on his device (e.g. \cite{collusion}). Moreover, most browsers allow to block third-party cookies (cookies that are set by a website not listed in the address bar, usually advertisements embedded on the primary website) \cite{ff-block,chrome-block}. 

This general unavailability of cookies motivated advertisers and trackers to find new ways to track users and their browsing histories. Today, there is a number of commercial
companies that provide ``cookieless" device identification through web-based fingerprinting~\cite{cookielessmonster}. In addition, there are numerous research works~\cite{mayer2009any, eckersley2010unique, yen2012host} that show that a device can be fingerprinted based on the browser and operating system features. The most notable features include plugin enumeration, enabled browser features, User-Agent HTTP header, font detection, Windows registry values, screen resolution and driver enumeration. For a comprehensive survey of the features used to fingerprint a device see~\cite{cookielessmonster}. All these features are solely software-based, namely they are features that can be set or changed using software only. As such they can be readily changed by the user. For example, by simply changing the browser used to access a website a user can foil many of these fingerprinting features. 

In this position paper we argue that the recent emergence of the HTML5 standard~\cite{html5} gives rise to a new class of fingerprinting features that are based on the \emph{hardware} of the device and are not necessarily dependent on the software installed on the device.  Such features are much harder to foil as long as the user uses the same device, hence they provide a stronger degree of confidence in the device fingerprint. These hardware features may be classified into two subclasses:

\begin{enumerate}
	\item Explicit -- these features are based on standard well-defined characteristics of the hardware usually determined explicitly by the manufacturer; most commonly, they allow to differentiate between different models or versions of the hardware module installed on the device. For example, the clock frequency of a GPU installed on the device is a standard explicitly designed characteristic of the model or version of the GPU.
	\item Implicit -- these features are derived from slight inconsistencies in the manufacturing process of the hardware module; they allow to differentiate between different instances of hardware having the \emph{same} model or version. For example, the clock skew of a GPU is different for every GPU instance even those that have the same clock frequency and model. 
\end{enumerate}

HTML5 offers a wealth of new features and capabilities to create more sophisticated and engaging web applications running over a browser. To allow these new features and capabilities the browser inevitably allows the web application a wider and more permissive access to the device's resources. In particular, HTML5 exposes high level JavaScript APIs that sit on top of the system's underlying hardware capabilities without the need to install third party plugins. This more permissive access to the device's hardware may allow a website to fingerprint a device based on its hardware features. Although very few works address hardware fingerprinting using HTML5, there are strong indications that suggests that this is feasible.

The structure of the paper is as follows: in Section~\ref{sec:relatedwork}  we review related work on device fingerprinting over the web. In Section~\ref{sec:hardwarehtml5} we introduce several possible hardware-based fingerprinting methods which may be used over the web. In Section~\ref{sec:gpu} we present results of initial experiments we conducted to fingerprint a device based on its GPU features. Section~\ref{sec:conclusions} concludes the paper.

\section{Related Work} \label{sec:relatedwork}
There are many works that explore methods to fingerprint a web client. As noted above, the vast majority of these works propose fingerprinting methods that are based on  software-related features, primarily those of the browser, operating system, and drivers. Ref.~\cite{eckersley2010unique} showed that parameters of system configuration such as screen resolution, browser plugins and system fonts as well as the contents of HTTP headers -- User-Agent and Accept -- allow to fingerprint a device. Other works proposed more fingerprinting methods that rely on other browser features, such as history and file cache~\cite{1624020}, JavaScript performance~\cite{moweryfingerprinting}, and deviations from JavaScript conformance tests~\cite{mullazanni13}. 

Additionally, in the past several years it has been shown~\cite{ayenson2011flash} that many web sites identify a web client based on ``super-cookies". These are identifiers which are stored on the local host in various persistent ways outside the control of a browser, hence the browser can not impose  standard restriction on their use as on HTTP cookies. This is also a software-based method.

To the best of our knowledge there are only two works that aim to fingerprint a web client based on its hardware features. Ref.~\cite{mowerypixel} proposed the use of rendering text and WebGL scenes to a HTML5 \textless canvas\textgreater~element while measuring the resulting pixel map of the canvas. Different browsers display text and graphics in a different way. The difference stems from a mix of software -- browser and driver -- and hardware -- GPU -- configurations. The proposed method does not intend to differentiate between two web clients with the exact same software and hardware configuration, i.e, using implicit hardware features.

Ref.~\cite{conf/aina/HuangYNTHL12} is the only work we are aware of that aim to differentiate between two web client based on implicit hardware features. The proposed approach is to measure the web client's CPU clock skew. The measurements are done using a JavaScript code that periodically sends timestamps back to the server. This work was inspired by  \cite{kohno2005remote} which showed that a device's clock skew can fingerprint a device.

\section{Hardware Access Using HTML5} \label{sec:hardwarehtml5}
In the following we list the hardware modules a web page may have direct to indirect access to using HTML5. For each device we reference the API documentation used to access that hardware module. And for each device we propose possible directions for a fingerprinting method.

\paragraph{GPU \cite{canvas}} All modern browser are now ``hardware-accelerated", meaning that rendering graphics may be done using the device's GPU. The rendering latency of various graphics may be used to fingerprint a GPU's clock frequency and skew.  In the next section we describe our initial experiments using this fingerprint method.

\paragraph{Camera \cite{camera-api}} A website may be able, by user permission, to capture a picture or a video using a camera. It has been shown in \cite{holst1998ccd} that each pixel of a camera's optical sensor has a small but measurable bias. This bias is a linear function of the actual intensity of light hitting the pixel\footnote{This linear bias is commonly called pattern noise.}. There are a few works that deal with camera identification using the pixels' bias. Ref.~\cite{camera_sensor} proposes a method to determine a camera's reference bias using 300 pictures taken by that camera. This serves as a unique fingerprint for the camera. It is shown that this enables to associate with good probability a new picture with the camera that took it.   

\paragraph{Speakers/Microphone \cite{audio-api}} The main specification of a microphone and speakers is the frequency response graph. A microphone's frequency response is its normalized output gain over a given frequency range. Conversely, a speaker's frequency response is its normalized output audio intensity over a given frequency range. A typical microphone or speaker has a response curve that varies across different frequencies. These variations are dependent on the design of the audio device. Moreover, due to manufacturing inconsistencies the frequency responses of each instance of a microphone or a speaker are not identical even if they are of the same model. A device's response for each frequency has a tolerance relative to the response specified by the manufacturer. A typical tolerance for commodity microphone and speakers is $\pm2$db. It has been suggested in \cite{sfgate-sensorid} that such variances in the frequency responses are able to fingerprint a mobile device. A web page can play tones in certain frequencies using the device's speakers while at the same time record the played audio using the microphone. This allows the web page to measure the frequency responses of the speakers and microphones. Such a method can also be applicable to desktop computers. 

\paragraph{Motion sensors \cite{motion-api}} Many mobile devices are equipped with accelerometer and gyroscope. Both sensors have linear biases, i.e. $v_m = v_t  S + O$. Here $v_m$ and $v_t$ denote the measured and true values, respectively, while $S$ and $O$ are the sensitivity and offset of the sensor. Ideally, the parameters' values should be $S=1$ and $O=0$. For each sensor instance the bias may be different due to manufacturing differences. Moreover, other imperfections, aside form the offset and sensitivity, may be created due to inconsistencies in the manufacturing process.  A web page can read measurements from these sensors, thereby potentially calculate their imperfections. Ref.~\cite{sfgate-sensorid} and \cite{AccelPrint} suggest that such a feat is indeed possible.
	
\paragraph{GPS \cite{GPS-api}} A GPS receiver triangulates the location of a device by calculating its distance to at least 3 GPS satellites. The distances are calculated by measuring the time a signal travels from a satellite to the GPS receiver. The travel time is measured using an inaccurate clock built into the GPS receiver. Previous work~\cite{kohno2005remote} has shown that a clock's skew can identify the clock. However, modern GPS receivers utilize a 4\textsuperscript{th} satellite measurement which allows to cancel out the bias. Therefore, theoretically, the clock's bias does not affect the calculated location. Nonetheless, there are still many sources of errors while calculating the receiver's position, such as atmospheric effects and multi-path effects. Such errors are not taken into account during the position determination and are implicitly treated as error sourced by the clock bias. Therefore, this may lead to a position calculation where the clock bias is no perfectly corrected. We hypothesize that taking multiple location measurements from the GPS might expose this bias.

\paragraph{Battery \cite{battery-api}} As time goes by the battery wears out and its capacity degrades. The rate at which this happens depends on many factors, such as climate, usage and charging patterns~\cite{spotnitz2003simulation}. The lose of capacity affects how fast a battery can be charged or discharged. Therefore, every device has a charging/discharging rate which is unique to it. A website can pole the battery status of a device (i.e. smartphone, tablet or laptop). Poling the battery allows to estimate the exact charging and discharging rate of a battery thereby fingerprinting the device. Of course, the charging/discharging rate is not dependent only on the battery capacity but also on the current activities performance on the device and the process that are currently running on it. A successful fingerprinting method would need to find a way to take away these factors.

\section{Fingerprinting Using the GPU} \label{sec:gpu}
In the following we present our initial efforts to fingerprint a device using its GPU. Although the results are encouraging, we do not aim to present here a complete GPU fingerprint method, but rather to present indications that such a method is possible. We aim to fingerprint the GPU based on an explicit feature -- the GPU clock frequency -- and based on an implicit feature -- the GPU clock skew. The latter feature is inspired by the clock skew method presented in \cite{kohno2005remote}. In this work a computer was fingerprinting using its CPU clock skew. In this method many timestamped packets were sent periodically by the device. Another computer picked up this packets and calculated the first computer's CPU's skew. More accurately, this method measured the first computer's CPU's skew relative to the CPU's skew of the second one. As one might expect, the network jitter between the two computers have a significant effect on the skew measurements. In order to average out this jitter many timestamped packet must by sent, thereby prolonging  the fingerprint process. Nonetheless, sending the timestamps to another computer is crucial to this fingerprinting method since one can not measure the skew of a clock without another reference clock. 

The basic idea of our fingerprinting method is simple. We leverage HTML5 in order to use the device's CPU as a reference clock while measuring the device's GPU clock skew. Namely, we strive to fingerprint a device by measuring the relative clock skews of the device's CPU and GPU. Unfortunately, unlike the CPU, there is no JavaScript API to get a timestamp of the GPU clock. Therefore, we have to resort to indirect methods in order to measure the GPU's clock skew. 

The fingerprinting method we employ is straightforward. We render using WebGL~\cite{webgl} complex 3D scenes on an HTML5 canvas. We measure how many frames are rendered on the screen within a predefined time interval. This measurements forms the basis for the GPU fingerprint. Assuming our fingerprint test is the only process that occupies the GPU then the number of rendered frames is largely dependent on explicit features of the GPU, e.g the clock frequency, number of cores and other parameters that affect the GPU's performance. Note, however, that the software that allows access to the GPU -- i.e. browser and GPU driver -- also influences to some extent the GPU performance.  The number of rendered frames are also affected, to a much lesser extent, by the clock skew of the GPU. Note that time interval within which the frames are rendered is measured using the CPU's clock.

Since the rate of rendered frames per second has a maximum value of 60 (this in the common refresh rate of computer displays), one must render complex graphics which strain the GPU so the frame rate will not reach this bound, thereby probing the GPU limits. 

\subsection{Experiment} 
In our experiment we crafted a website that displays 3D graphics in three phases. Each phase displayed increasingly complex graphics. Where each phase lasted 15 seconds. within which we measured the number of rendered frames in three 5-seconds intervals. The rendering process was performed using the requestAnimationFrame() API. This API signals the browser that the animation procedure should be called before the next refresh of the device's display. The animation procedure itself calls this API so it shall be invoked regularly. The experiment website can be found here: \url{http://fingerprintme.herobo.com}. We calculate the average of the 3 measurements of each phase. Thereby getting a fingerprint per phase for each device. We fingerprinted 130 different devices; 34 of which were fingerprinted more than once. To simplify the analysis we confined our experiment only to desktop and laptop computers. 

\begin{figure*}[t]
\begin{center}
	\subfigure[phase I results (least complex graphics)]{
		\label{fig:results:a} %% label for first subfigure
		\includegraphics[width=0.31\textwidth]{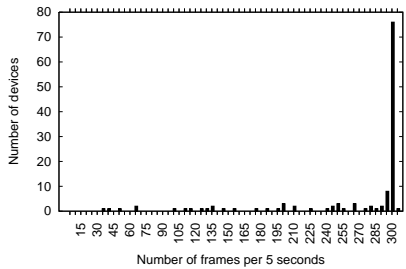}}
	\hspace{0.0cm}
	\subfigure[phase II results (medium complexity graphics)]{
		\label{fig:results:b} %% label for second subfigure
		\includegraphics[width=0.31\textwidth]{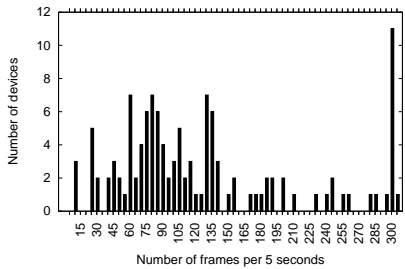}}
	\hspace{0.0cm}
	\subfigure[phase III results (most complex graphics)]{
		\label{fig:results:c} %% label for second subfigure
		\includegraphics[width=0.31\textwidth]{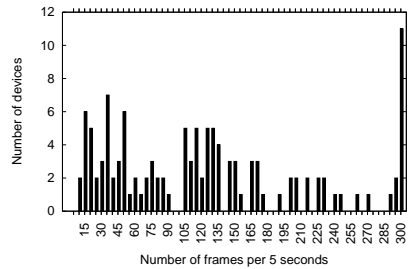}}
	\caption{Distribution of number of rendered frames}
	\label{fig:results} %% label for entire figure
	\end{center}
\end{figure*}

\subsection{Results} 
We first measured for each computer and each phase the 95\textsuperscript{th} percentile of the difference between every two measurements (i.e. the number of rendered frames per 5 seconds).  We used the 95\textsuperscript{th} percentile rather than the maximum difference to disregard any outliers which are most notably caused by the browser window leaving focus and not showing on the screen\footnote{In such cases the browser automatically reduces the maximum number of rendered frames per second.}. The 95\textsuperscript{th} percentile difference was calculated to be 5 frames. We then plotted for each phase a histogram of the phase's measurements with bins having width of 5 frames. The histograms are depicted in Figure~\ref{fig:results}. The entropy of the fingerprint distribution for each phase can be found in Table~\ref{tbl:results}.

It is evident from both Figure~\ref{fig:results} and Table~\ref{tbl:results} that as the rendered graphics are more complex the fingerprint allows to better differentiate between different computers. The most complex phase is able to yield fingerprint entropy of 5.14 bits. We suspect that even more complex graphics would yield higher entropy. 

Note that this initial experiment actually measures the performance of the GPU. As noted above, although the most influential features of this performance is the hardware features of the GPU. The efficiency of the software that uses the GPU (browser and GPU driver) also affect its performance. Therefore, more sophisticated fingerprint tests may be needed to weed out the software influence.

\begin{table}
	\centering
		\begin{tabular}{|c|c|c|}
		\hline
		Phase I & Phase II & Phase III \\
		\hline
		2.8 & 5.03 & 5.14\\
		\hline
		\end{tabular}
		\caption{The entropy of the fingerprint for each phase [bits]} \label{tbl:results}
\end{table}

We also run the fingerprint test on 9 desktop computers having the same hardware and software configuration\footnote{computer model: HPZ220, graphics card: Nvidia Quadro 600, OS: Windows 7 SP1, Browser: Firefox.}.The results of the fingerprints are in Table~\ref{tbl:samespec}. The fingerprint was consistent across multiple experiments for each computer. The results give encouraging indication that a GPU fingerprint may allow to differentiate even computers with the same hardware and software configurations. The differences in fingerprints were actually wider than we initially expected. It seems that more factors, other than the GPU clock skew, differ between GPUs of the same model. A successful fingerprinting method would need to take this into account. 

\begin{table}
	\centering
		\begin{tabular}{|c|c|c|}
		\hline
		200 & 236 & 234 \\
		\hline
		249 & 177 & 195 \\
		\hline
		182 & 190 & 232 \\
		\hline
		\end{tabular}
		\caption{Fingerprint value measured during phase III for 9 different computers having the same hardware and software specification} \label{tbl:samespec}
\end{table} 

\section{Conclusion} \label{sec:conclusions}
In this paper we point out the possibility that the recent emergence of HTML5 enable a new class of powerful fingerprinting features that are based on the characteristics of the hardware of the device rather than on the software it runs. Such hardware features allow not only to differentiate between devices with different hardware configurations, but also can potentially differentiate between devices having the same exact hardware configuration. We introduce some ideas for hardware fingerprinting. And present encouraging results of an initial experiment to fingerprint a device using its GPU.  

\bibliographystyle{IEEETran}
\bibliography{GPU-ID}

\end{document}